\documentclass[twocolumn,showpacs,showkeys,preprintnumbers,amsmath,amssymb]{revtex4}
\usepackage{epsfig,amsfonts}
\usepackage{amsmath}
\usepackage{bbm}
\usepackage{wasysym}
\usepackage{graphicx,psfrag,rotating}
\usepackage{graphics}
\usepackage{txfonts}

\usepackage{textcomp}
\usepackage{graphicx}% Include figure files
\usepackage{dcolumn}% Align table columns on decimal point
\usepackage{bm}% bold math

\begin{document}

%\preprint{APS/123-QED}%

\title{The properties of C-parameter and coupling constants}

\author{R. Saleh-Moghaddam and M.E. Zomorrodian}
\affiliation{Department of Physics, Faculty of physics, Ferdowsi University of Mashhad, 91775-1436 Mashhad, Iran\\ $R‎_{-}‎$Saleh88@yahoo.com , Zomorrod@um.ac.ir}

\date{March 10, 2015}

\begin{abstract}
‎\textbf{abstract:}‎ In this article, we present the properties of the C-parameter which is one of event shape variables. We obtain the coupling constants both in the perturbative and in the non-perturbative part of the QCD theory. To achieve this we fit the dispersive model as well as the shape function model with our data. Our results are consistent with the QCD predictions. We explain more features of our results in the main text. 
\end{abstract}
\pacs{12.38.-t; 12.38.Bx; 12.38.Lg}
\keywords{Quantum chromo dynamics; Perturbative calculations; non-Perturbative theory}

\maketitle

\section{Introduction}
Event shape variables in $e^{+}e^{-}$ annihilation provide an ideal testing ground to study Quantum Chromo-Dynamics (QCD) and have been measured and studied extensively in the last three decades. In particular, event shape variables are interesting for studying the interplay between perturbative and non-perturbative dynamics \cite{ref1}. One of the most common and successful ways of testing QCD has been by investigating the distribution of event shapes in $e^{+}e^{-}\rightarrow hadrons$, which have been measured accurately over a range of centre-of-mass energies, and provide a useful way of evaluating the strong coupling constant $\alpha_{s}$. The main obstruction to obtaining an accurate value of $\alpha_{s}$  from distributions is not due to a lack of precise data but to dominant errors in the theoretical calculation of the distributions. In particular, there are non-perturbative effects that cannot yet be calculated from first principles but cause power-suppressed corrections that can be signiﬁcant at experimentally accessible energy scales \cite{ref2}. 

In this article, we show the cross section for C- parameter as an event shape observable. Also we peruse both perturbative and non-perturbative theory for calculating coupling constants using different models .It is mentioned that we have already done some analyses on some event shape observables in our previous publications \cite{ref3, ref4}.
 
 The outline of the paper is as the following: In sect.2 we define and review the C-parameter and show the cross section distribution for different energies. In sect.3, we present the calculations of the perturbative theory as well as the non-perturbative theory up to the next to next to leading order (NNLO). We also achieve similar calculations for both the dispersive and the shape function models. And finally we extract the coupling constant from our analysis. The last section summarizes our conclusions.

\section{C-parameter}
The C parameter \cite{ref5, ref6} for electron$-$positron annihilation events is derived from the eigenvalues $\lambda_{i}$ of the linearized momentum tensor $\theta_{jk}$. 
\begin{equation}\label{1} 
\theta_{jk}=\dfrac{\sum_{i}p^{i}_{j}p^{i}_{k}/\vert p^{i}\vert}{\sum_{i}\vert p^{i}\vert}.
\end{equation}
where $p^{i}$ are the spatial components $( j, k= 1, 2, 3)$ of the \textit{i-th} particle momentum in the centre of mass frame. The sum on \textit{i} runs over all the final state particles. If $\lambda_{i}$ are the eigenvalues of the matrix, we have:
\begin{equation}\label{2}
C=3(\lambda_{1}\lambda_{2}+\lambda_{2}\lambda_{3}+\lambda_{3}\lambda_{1}).
\end{equation}

The real symmetric matrix $\theta_{jk}$ has eigenvalues $\lambda_{i}$ with $ 0\leq\lambda_{3}\leq\lambda_{2}\leq\lambda_{1}\leq1$. It describes an ellipsoid with orthogonal axes named minor, semi-major and major corresponding to the three eigenvalues. The major axis is similar but not identical to $n_{T}$. The C parameter varies in the range $0‎\leq‎C‎\leq‎1$. $C=0$ corresponds to a perfect two jet event (with massless jets), while $C=1$ characterizes a spherical event. Planar events including in particular the $O(\alpha_{S})$ perturbative result, are distributed in the range $0‎\leq‎C‎‎\leq‎‎3/4$.

\begin{figure}[h!]
\centering
\includegraphics[scale=.3]{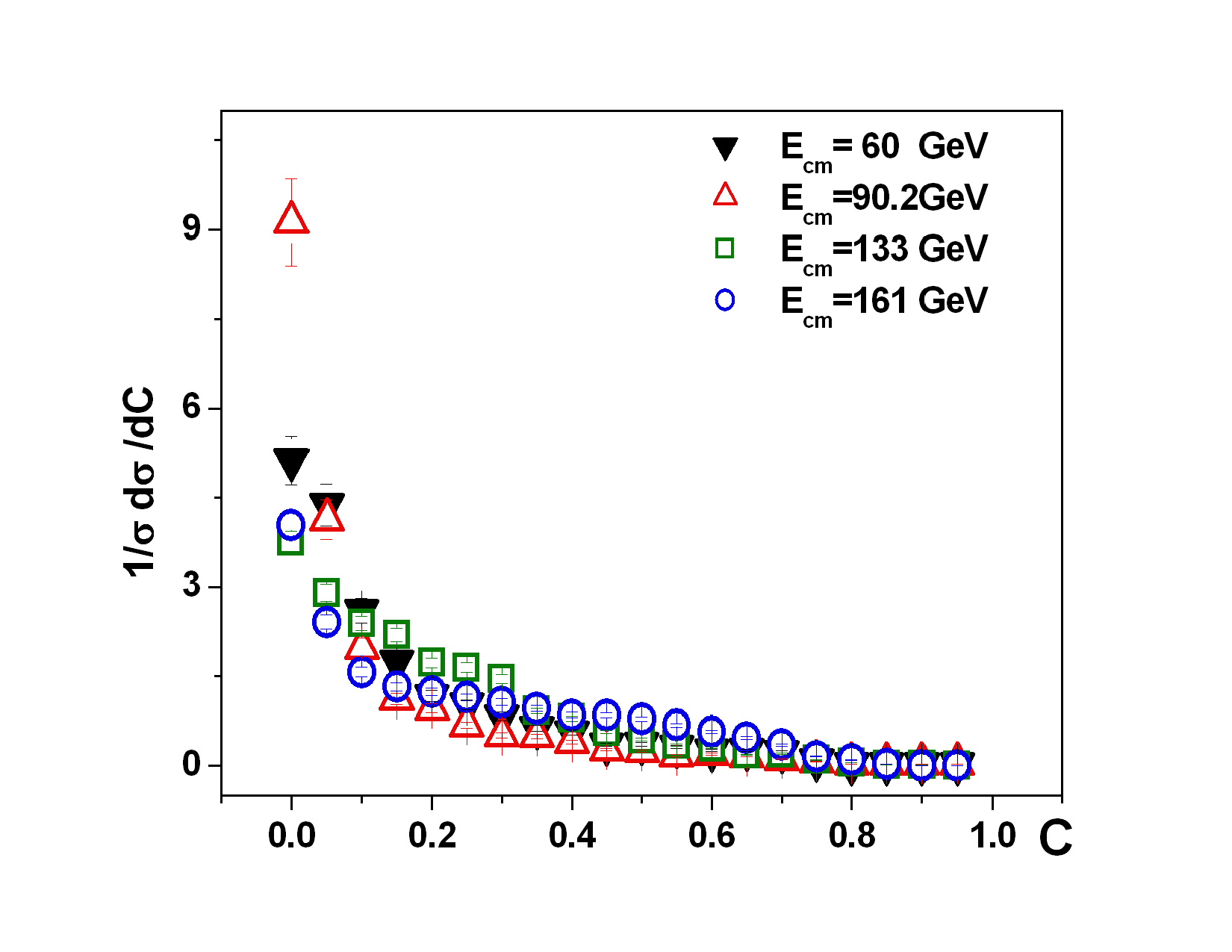}
\caption{The cross section distributions for C-parameter in different energies.}
\label{fig1}
\end{figure}

Figure \ref{fig1} shows the distribution for C-parameter $(0‎\leq‎C‎\leq‎1)$ by using the Monte Carlo data (PYTHIA program) in different center of mass energies. We see that the values are between 0 and 1 for all energies. We observe a decreasing trend for all distributions by increasing the C parameter. The figure also indicates that there is not a significant difference on the distributions between different energies. We conclude that this parameter is insensitive to the data of different energies.

‎\section{Power corrections and different models}‎
QCD calculations based on NLO perturbation theory as well as NNLO theory are available for many observables in high energy particle reactions, like the total hadronic cross section in $e^{+} e^{-}\rightarrow hadrons$ and moments in deep inelastic scattering (DIS) processes and also in  hadronic decay. The complicated nature of QCD, due to the process of gluon self coupling and the resulting large number of Feynman diagrams in higher orders of perturbation theory, so far limited the number of QCD calculations in complete NNLO \cite{ref7}. We shall explain the methods of calculation for NLO and NNLO theories. We also give an explanation of other ways for calculating the coupling constant such as the dispersive model and also the shape function model. These models include perturbative and non-perturbative regions. One of the aspects where the study of event shapes has taught us general lessons about QCD is the interplay between perturbative and non–perturbative corrections. The phenomenological success of renormalization based on models for power corrections has important consequences.

In event-shape moments \textit{(y)}, one expects the hadronization corrections to be additive, such that they can be divided into a perturbative and a non-perturbative contribution \cite{ref8},

\begin{equation}\label{3}
\langle y^{n}\rangle = \langle y^{n} \rangle_{pt} + \langle y^{n} \rangle_{np} .
\end{equation}

Now we explicate these theories and models in the following:

\subsection{\label{3-1}NLO and NNLO corrections}
The perturbative expansion of a differential distribution weighted of the observable y can be written for any infrared-safe observable for the process $e^{+} e^{-}\rightarrow hadrons$. The main theoretical feature of jet observables is that they can be computed in QCD perturbation theory. As a result, jets observables are finite (calculable) at the partonic level order by order in perturbation theory \cite{ref7}. If we assume the NLO theory only, we can compute its corresponding perturbative expansion in the form 
\begin{equation}\label{4}
\dfrac{1}{\sigma_{tot}}\dfrac{d\sigma}{dy}=(\dfrac{\alpha_{s}(\mu)}{2\pi}) \ \dfrac{d\overline{A}}{dy}
+(\dfrac{\alpha_{s}(\mu)}{2\pi})^{2} \ \dfrac{d\overline{B}}{dy}.
\end{equation}
$\overline{A}$ gives the LO result and $\overline{B}$ the NLO correction. $\sigma_{tot}$ denotes the total hadronic cross section calculated up to the relevant order. The arbitrary renormalization scale is denoted by $\mu$. All observables are chosen such that they take values between 0 and 1. 
Now if we add the third sentence to above expansion, we will have the NNLO theory.
\begin{equation}\label{5}
\dfrac{1}{\sigma_{tot}}\dfrac{d\sigma}{dy}=(\dfrac{\alpha_{s}(\mu)}{2\pi}) \ \dfrac{d\overline{A}}{dy}+(\dfrac{\alpha_{s}(\mu)}{2\pi})^{2} \ \dfrac{d\overline{B}}{dy}
+(\dfrac{\alpha_{s}(\mu)}{2\pi})^{3} \ \dfrac{d\overline{C}}{dy}.
\end{equation}
$\overline{C}$ gives the NNLO correction. The perturbative contribution to $\langle y^{n}\rangle$  is given up to NNLO in terms of the dimensionless coefficients $\overline{A}_{y,n}$, $\overline{B}_{y,n}$  and $\overline{C}_{y,n}$  as \cite{ref1}:
\[\langle y^{n}\rangle_{PT} = (\dfrac{\alpha_{s}(\mu)}{2\pi})\overline{A}_{y,n}+(\dfrac{\alpha_{s}(\mu)}{2\pi})^{2}\]
\[(\overline{B}_{y,n}+\overline{A}_{y,n} \beta_{0} log\dfrac{\mu^{2}}{E_{cm}})+(\dfrac{\alpha_{s}(\mu)}{2\pi})^{3}(\overline{C}_{y,n}+2\overline{B}_{y,n} \beta_{0}\]
\begin{equation}\label{6}
 log\dfrac{\mu^{2}}{E_{cm}}+\overline{A}_{y,n}(\beta_{0}^{2} log^{2}\dfrac{\mu^{2}}{E_{cm}}+\beta_{1} log\dfrac{\mu^{2}}{E_{cm}}).
\end{equation}
$A_{y,n}$, $B_{y,n}$ and $C_{y,n}$ are straightforwardly related to $\overline{A}_{y,n}$, $\overline{B}_{y,n}$ and $\overline{C}_{y,n}$:
\[\overline{A}_{y,n}=A_{y,n},\]
\[\overline{B}_{y,n}=B_{y,n}-\frac{3}{2} C_{F} A_{y,n},\]
\begin{equation}\label{7}
\overline{C}_{y,n}=C_{y,n}-\frac{3}{2} C_{F} B_{y,n}+(\frac{9}{4} C_{F}^{2}-K_{2})A_{y,n}.
\end{equation}
that $A_{y,n}$, $B_{y,n}$ and $C_{y,n}$ are given in Table ‎\ref{tab1}‎ also the constant $K_{2}$ is given by \cite{ref9, ref10}:
\[K_{2}=\frac{1}{4}[-\frac{3}{2}C_{F}^{2}+C_{F}C_{A}(\frac{123}{2}-44\zeta_{3})\]
\begin{equation}\label{8}
+C_{F}T_{R}N_{F} (-22+16\zeta_{3})].
\end{equation}

\begin{table}[b]\caption{\label{tab1}
Contributions to C-parameter at LO, NLO and NNLO \cite{ref1}.}
\centering
\begin{tabular}{||c|c|c|c||}\hline 
\textrm{\textbf{$\langle C^{n} \rangle$}}&\textrm{\textbf{A}}&\textrm{\textbf{B}}&\textrm{\textbf{C}}\\
\hline
n=1& 8.6379 & $172.778\pm 0.007$ & $3212.2\pm 88.7$ \\
n=2 &2.4317 & $81.184\pm 0.005$ & $2220.9\pm 12.0$ \\ 
n=3 & 1.0792 & $42.771\pm 0.003$ & $1296.6\pm 6.7$ \\ 
n=4 & 0.5685 & $25.816\pm 0.002$ & $843.1\pm 3.9$ \\
n=5 & 0.3272 & $16.873\pm 0.001$ & $585.0\pm 3.2$ \\ \hline
\end{tabular}
\end{table}
Here $E_{cm}$ denotes the centre of mass energy squared and μ is the QCD renormalization scale. The NLO expression is obtained by suppressing all terms at order $\alpha_{S}^{3}$. The first two coefficients of the QCD $\beta$-function are:
\[\beta_{0}=\frac{11C_{A}-4T_{R}N_{F}}{6},\]
\begin{equation}\label{9}
 \beta_{1}=\frac{17C_{A}^{2}-10C_{A}T_{R}N_{F}-6C_{F}T_{R}N_{F}}{6}.
\end{equation}
where $C_{A}=N$ is color number, $C_{F}=(N^{2}-1)/2N=4/3$, $T_{R}=1/2$ and $N_{F}$ is quark flavors.

The perturbative coefficients ($\overline{A}_{y,n}$, $\overline{B}_{y,n}$ and $\overline{C}_{y,n}$) are independent of the centre-of-mass energy. They are obtained by integrating parton-level distributions, which were calculated recently to NNLO accuracy \cite{ref1}. These values are shown in Table \ref{tab1}.

In order to define the non-perturbative part of the theory, it is necessary to explain the dispersive as well as the shape function model as the following.

\subsection{\label{3-2}The dispersive model}
Non-perturbative power corrections can be related to infrared renormalizations in the perturbative QCD expansion for the event-shape variables. The dispersive model for the strong coupling leads to a shift in the distributions \cite{ref8}:
\begin{equation}\label{10}
\frac{d\sigma}{dy}(y)=\frac{d\sigma_{pt}}{dy} (y-a_{y}.P)
\end{equation}
where the numerical factor $\alpha_{y}$ depends on the event shape that  for C-parameter it is $a_{y}=3‎\pi‎$ \cite{ref11}.
Then we obtain the non-perturbative parameter from this model as the following: 

\begin{equation}\label{11}\nonumber
\langle y^{n} \rangle_{np}=a_{y}.P=a_{y}.\frac{4C_{F}}{\pi^{2}}\textit{M} \frac{\mu_{I}}{E_{cm}}
\end{equation}
\begin{equation}
.[\alpha_{0}(\mu_{I})-\alpha_{s}(\mu)-(log(\frac{\mu}{\mu_{I}})+1+\frac{k}{4\pi\beta_{0}}) \ 2 \ \beta_{0} \ \alpha_{s}^{2}(\mu)]
\end{equation}

In the $\overline{MS}$ renormalization scheme the constant $\textit{k}$ has the value  $k=(67/18-\pi^{2}/6) C_{A}-5/9 N_{F}$ , The Milan Factor $\textit{M}$ is known in two loops as $\textit{M}=1.49\pm0.20$ \cite{ref11, ref12}, for number of flavor $N_{F}=3$ at the relevant low scales \cite{ref13}. 

As a result for $\textit{M}$ that so-called non-inclusive Milan factor, we had \cite{ref14}: 

\begin{equation}\label{12}\nonumber
\textit{M}=1+\frac{3.299C_{A}}{\beta_{0}}+2\times\frac{-0.862C_{A}-0.052N_{F}}{\beta_{0}}
\end{equation}
\begin{equation}
=1+\frac{1.575C_{A}-0.104N_{F}}{\beta_{0}}=1.49\pm 0.20
\end{equation}

Now we can use the power corrections to calculate the perturbative and the non-perturbative theories for the moments of y, we have \cite{ref8}:

\begin{equation}\label{13}
\langle y^{1} \rangle = \langle y^{1}\rangle_{NLO}+a_{y}.P,
\end{equation}
\begin{equation}\label{14}
\langle y^{2} \rangle = \langle y^{2}\rangle_{NLO}+2\langle y^{1}\rangle _{NLO}(a_{y}.P)+(a_{y}.P)^{2},
\end{equation} 
\begin{equation}\label{15}
\langle y^{3} \rangle = \langle y^{3}\rangle_{NLO}+3\langle y^{2}\rangle _{NLO}(a_{y}.P)
+3\langle y^{1}\rangle _{NLO}(a_{y}.P)^{2}+(a_{y}.P)^{3}.
\end{equation}

\begin{figure}%The best place to locate the table environment is directly after its first reference in text%
\centering
\includegraphics[scale=0.23]{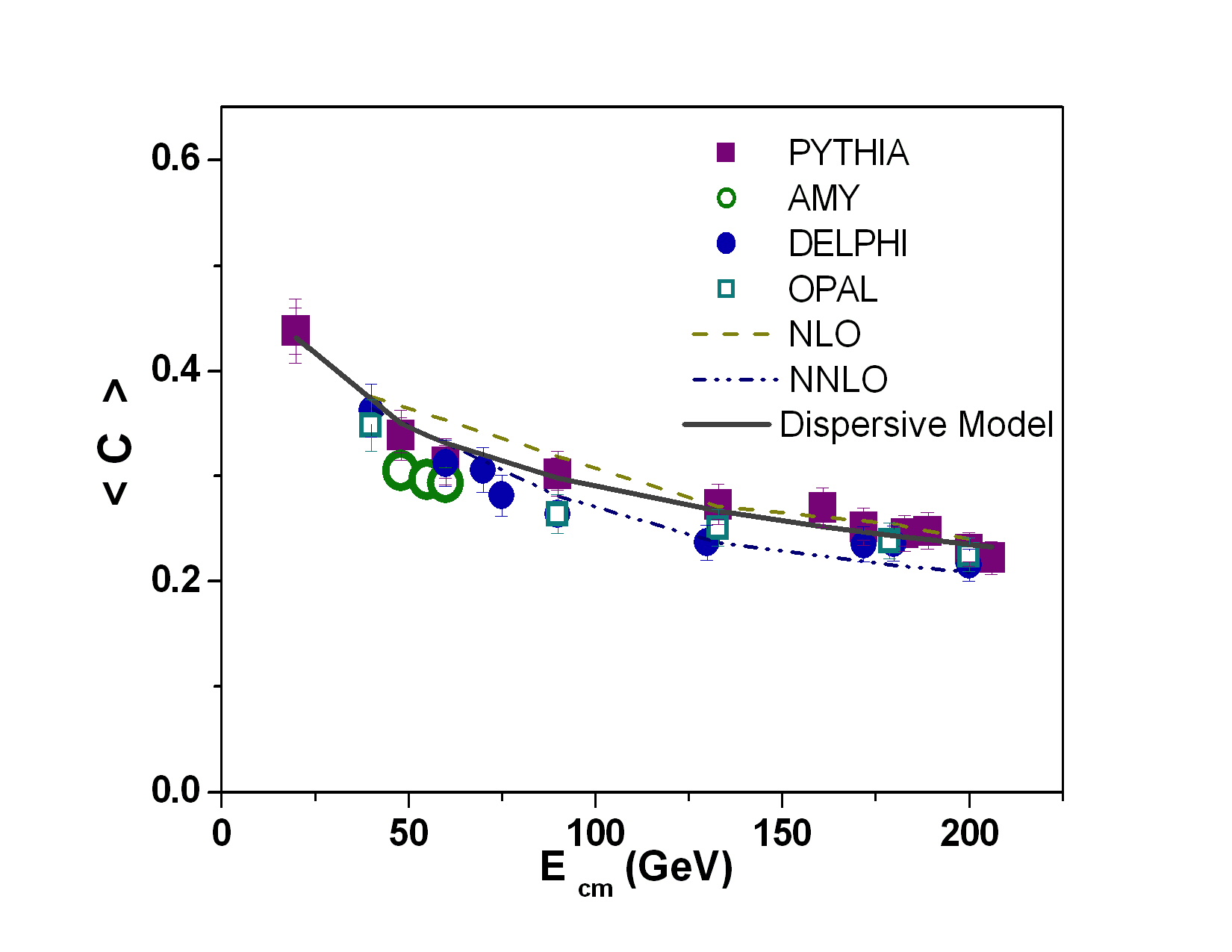}
\includegraphics[scale=0.23]{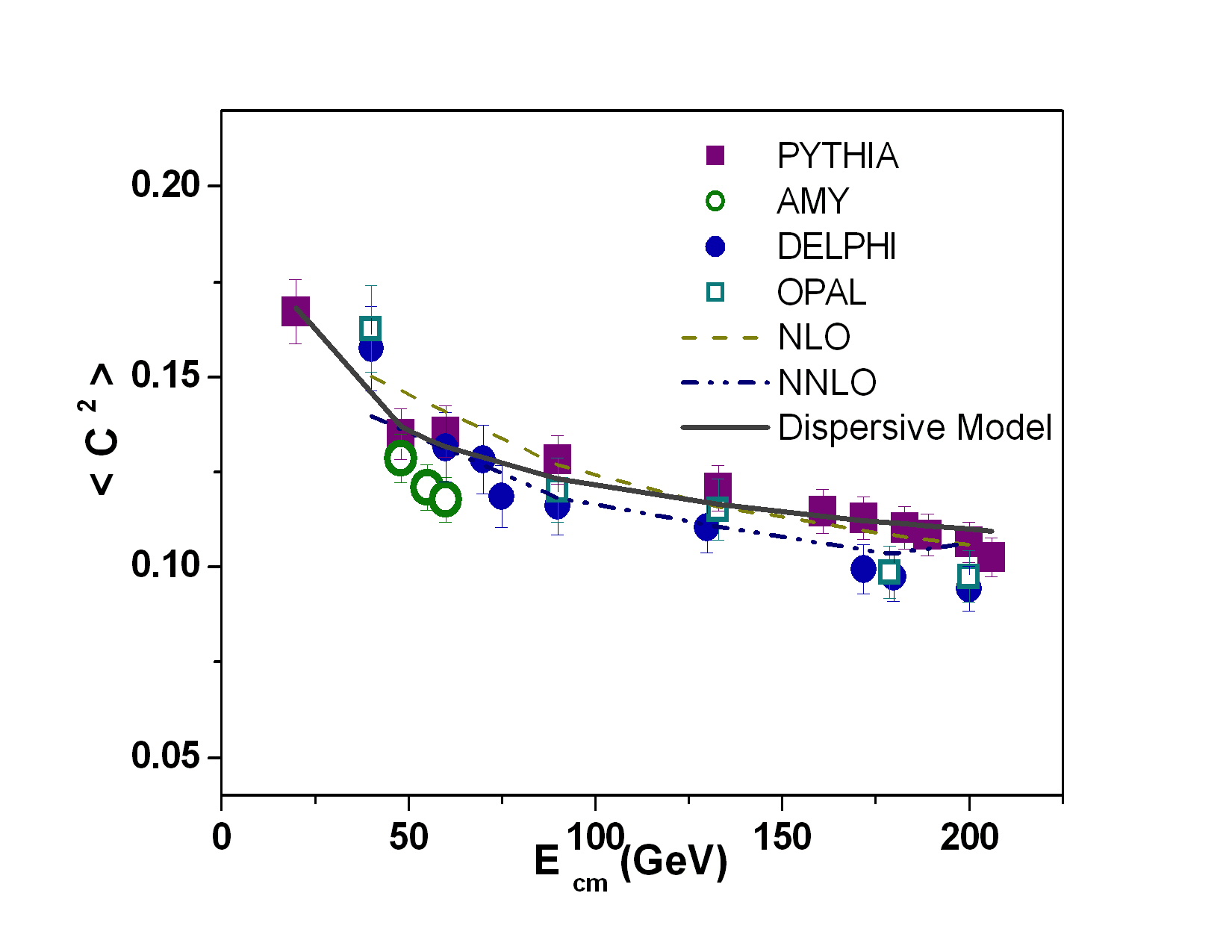}
\includegraphics[scale=0.23]{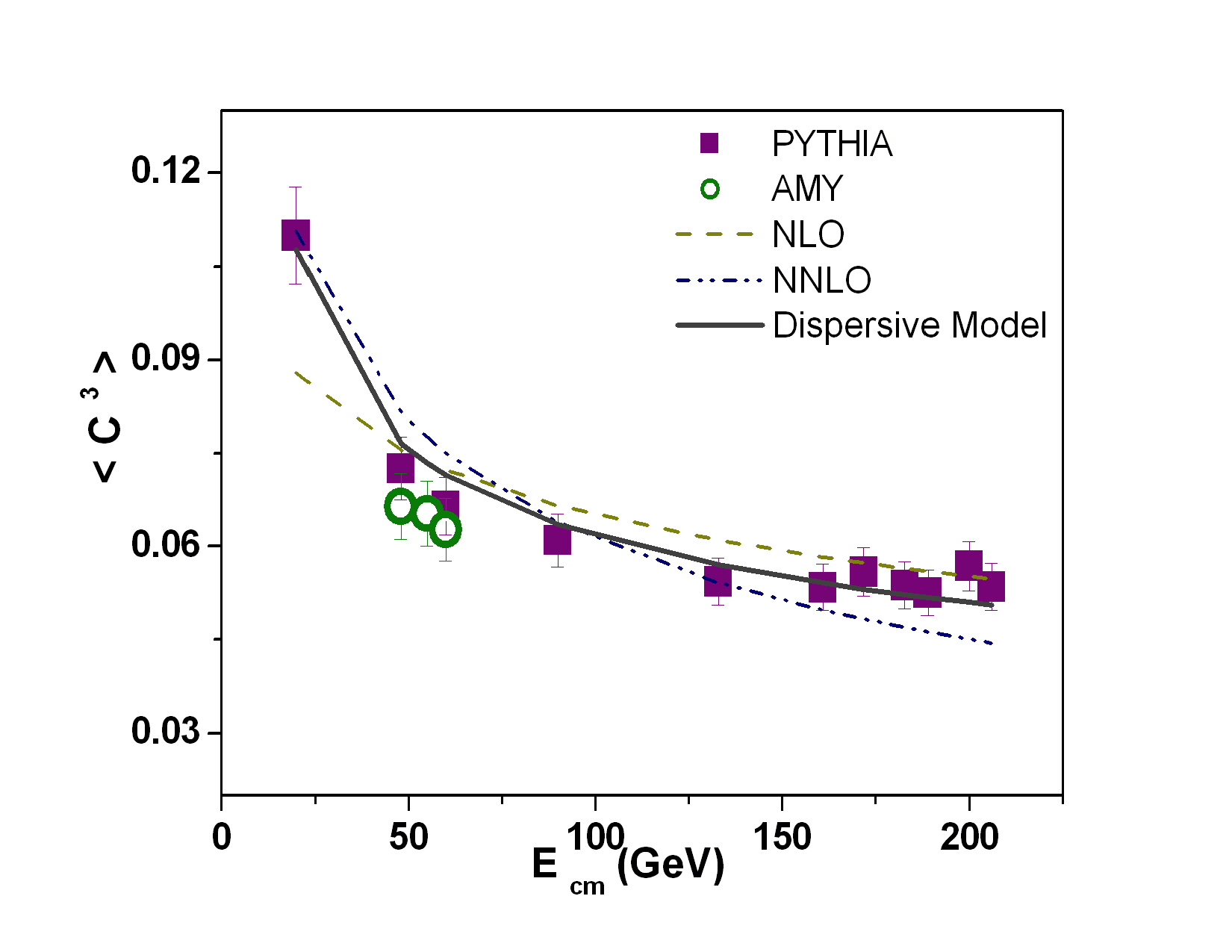}
\caption{\label{fig2} Fitting the dispersive model with data for the C-parameter}
\end{figure}

$\alpha_{s}$ defines the strong coupling constant and $\alpha_{0}$ defines the nonperturbative parameter accounting for the contributions to the event shape below an infrared matching scale $\mu_{I}\cong2$. Therefore it is possible to obtain $\alpha_{S}$ and $\alpha_{0}$ in this model. We are using the AMY data taken from TRISTAN at KEK, as well as DELPHI and ALEPH at CERN plus the PYTHIA event generator. By fitting the dispersive model with diagrams depicted in Figure \ref{fig2}, we observe that the model is consistent with the data. 

We observe that all distributions for the dispersive model (solid line) are in good agreement with the Monte Carlo (PYTHIA) and the experimental data when compared with NLO or NNLO prediction because the latters do not include the power correction. In other words the dispersive model includes both perturbative and the non-perturbative part of the theory. Next step is to measure the strong coupling constant and the non- perturbative parameter by fitting the above model with the data. The values for $\alpha_{s} (M_{Z^{0}})$ and $\alpha_{0} (\mu_{I})$ up to third power correction are cited in Table \ref{tab2}. Our results are also consistent with those obtained by using other experiments (Table \ref{tab3} is extracted from \cite{ref15}).

\begin{table}[b]%The best place to locate the table environment is directly after its first reference in text
\caption{\label{tab2}%
Measurements of the coupling constants using the dispersive model.}
\centering
\begin{tabular}{||c|c|c|c||}\hline
\textrm{\textbf{n}}&
\textrm{\textbf{1}}&
\textrm{\textbf{2}}&
\textrm{\textbf{3}}\\ \hline
\textbf{$\alpha_{s}(M_{Z^{0}})$} & $0.149\pm 0.003$ & $0.117\pm 0.001$ &$ 0.112\pm 0.003$ \\
\textbf{$\alpha_{0}(\mu_{I})$} & $0.460\pm 0.015$ & $0.471 \pm 0.007$ & $0.499\pm 0.017$ \\ \hline
\end{tabular}

\end{table}

\begin{table}[b]%The best place to locate the table environment is directly after its first reference in text
\caption{\label{tab3}%
Coupling constants for C-parameter in \cite{ref15}.}
\centering
\begin{tabular}{||c|c|c|c||}\hline
\textrm{\textbf{Exp}}&
\textrm{\textbf{ALEPH}}&
\textrm{\textbf{DELPHI}}&
\textrm{\textbf{L3}}\\
\hline
\textbf{$\alpha_{s}(M_{Z^{0}})$} & $0.123\pm 0.003$ & $0.122\pm 0.004$ & $0.116\pm 0.005$ \\
\textbf{$\alpha_{0}(\mu_{I})$} & $0.461\pm 0.016$ & $0.444\pm 0.022$ & $0.457\pm 0.040$ \\  \hline
\end{tabular}
\end{table}
\
\subsection{\label{3-3}The shape function model}
Korchemsky and Tafat \cite{ref16} describe properties of the event shape variables $1-T$ and $M_{H}^{2}$ not included in NLO perturbation theory so called shape function, which does not depend on the variable nor the cms energy. This is more general than the dispersive model, considering both as a shift of the perturbative prediction and as a compression of the distribution peak.  We should mention that the shape function model includes the perturbative theory as well as the non-perturbative part of theory. This model is a combination of both the NLO prediction and the power correction terms (equation \ref{3}). To calculate the strong coupling constant in perturbative theory, we are using equation \ref{4}. We also use the following expansion for measuring the free parameter in non-perturbative theory.   
\begin{equation}\label{16}
\langle C^{1} \rangle = \langle C^{1}\rangle_{NLO}+\frac{\lambda_{1}}{E_{cm}}.
\end{equation}

Analogously for the second moments and the third moments, we find \cite{ref17}:
 \begin{equation}\label{17}
\langle C^{2} \rangle = \langle C^{2}\rangle_{NLO}+2\frac{\lambda_{1}}{E_{cm}}\langle C^{1}\rangle_{NLO}+
\frac{\lambda_{2}}{E_{cm}^{2}}.
\end{equation}
\begin{equation}\label{18}
\langle C^{3} \rangle = \langle C^{3}\rangle_{NLO}+3\frac{\lambda_{1}}{E_{cm}}\langle C^{2}\rangle_{NLO}+3\frac{\lambda_{2}}{E_{cm}^{2}}\langle C^{1}\rangle_{NLO}+\frac{\lambda_{3}}{E_{cm}^{3}}.
\end{equation}

In order to calculate the strong coupling constant concerned with the perturbative theory, we perform a fitting procedure by including just the first part of equation \ref{16}. On the other hand to find the coupling constant in the non-perturbative region we do a fitting procedure by taking into account both parameters in equation \ref{16}. We achieve a similar analysis for equations \ref{17} and \ref{18}. The coefficient $\lambda_{1}$ is interpreted as the first “moment” and $\lambda_{2}$ as the second moment of the shape function. These parameters are also known as the so called universal scales \cite{ref18}.

\begin{figure}
 \centering
\includegraphics[scale=0.23]{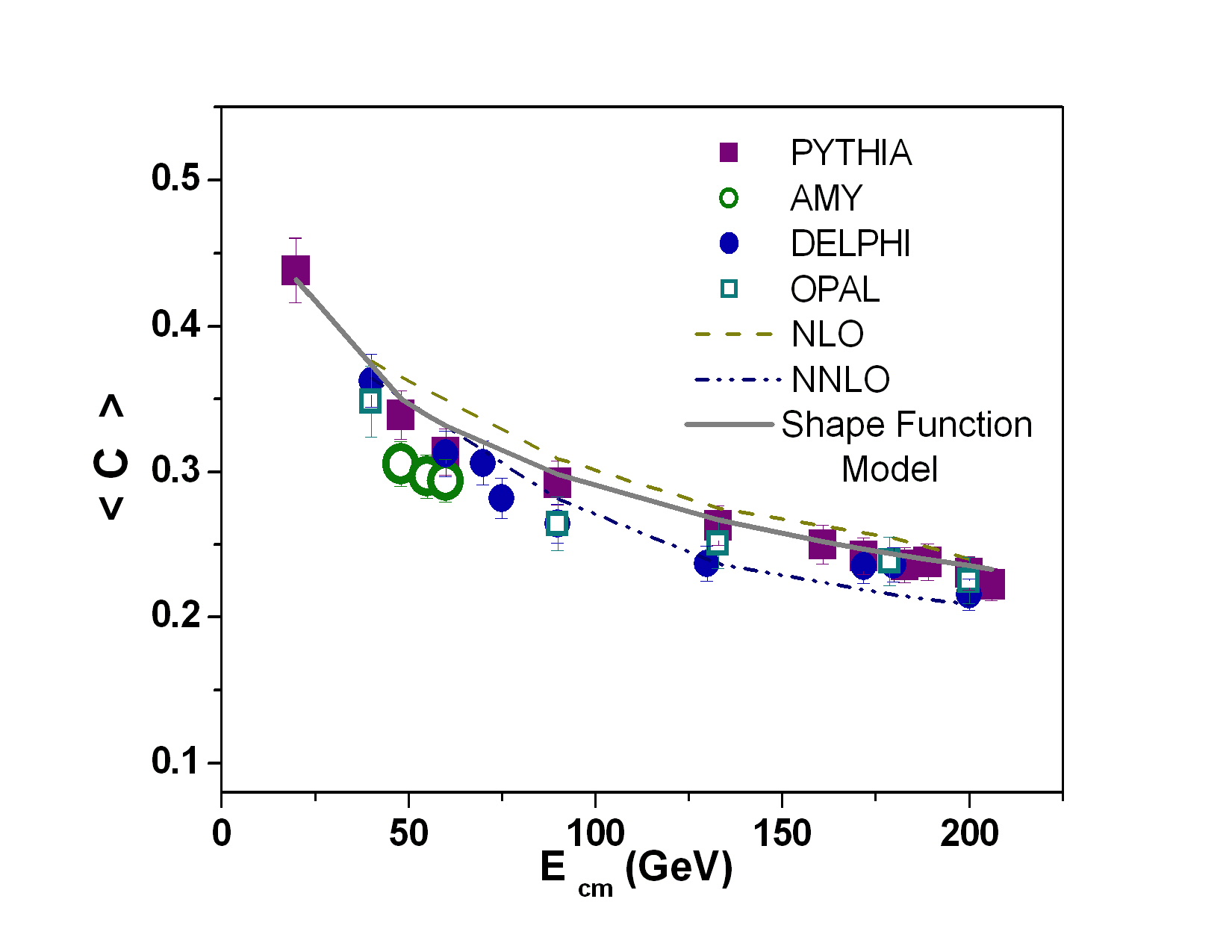}
\includegraphics[scale=0.23]{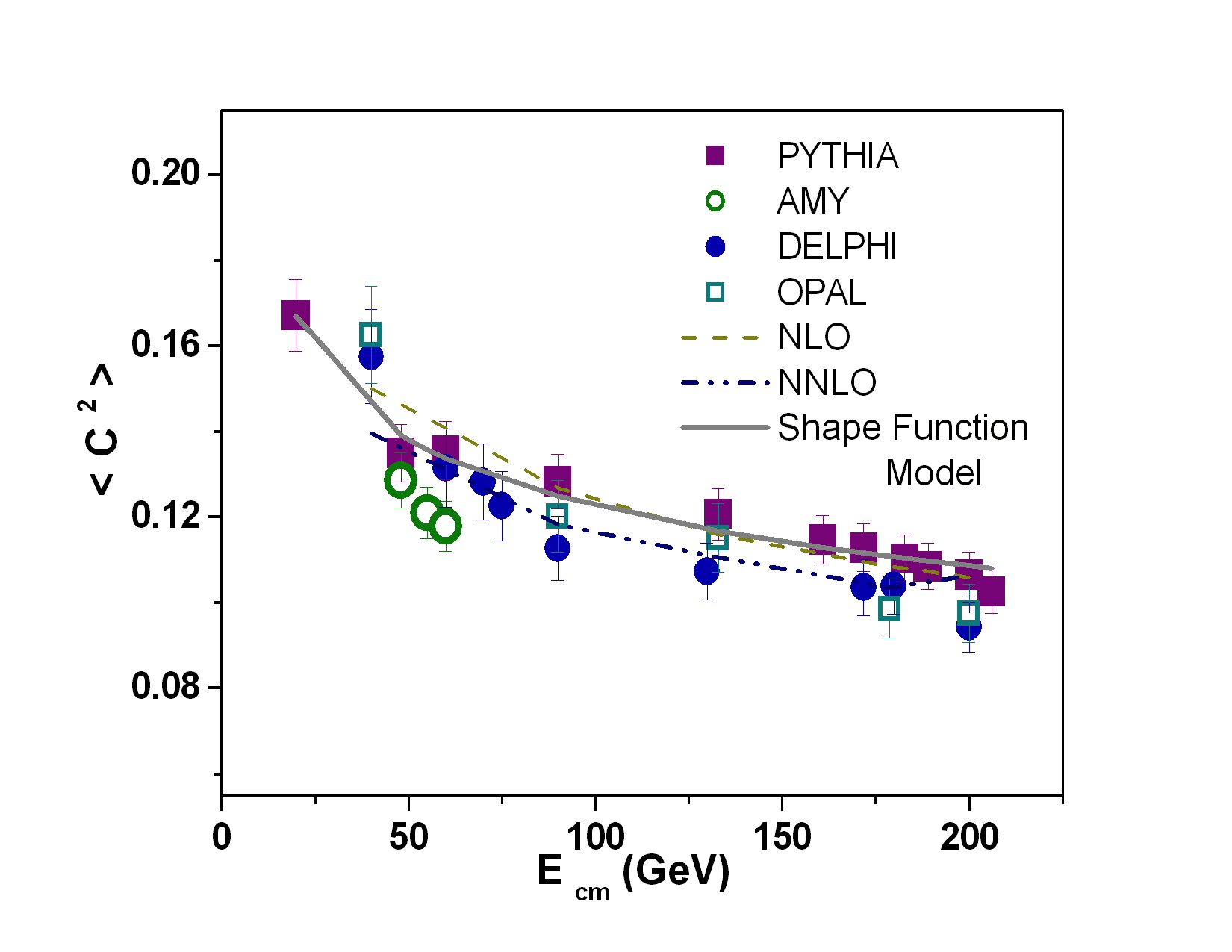}
\includegraphics[scale=0.23]{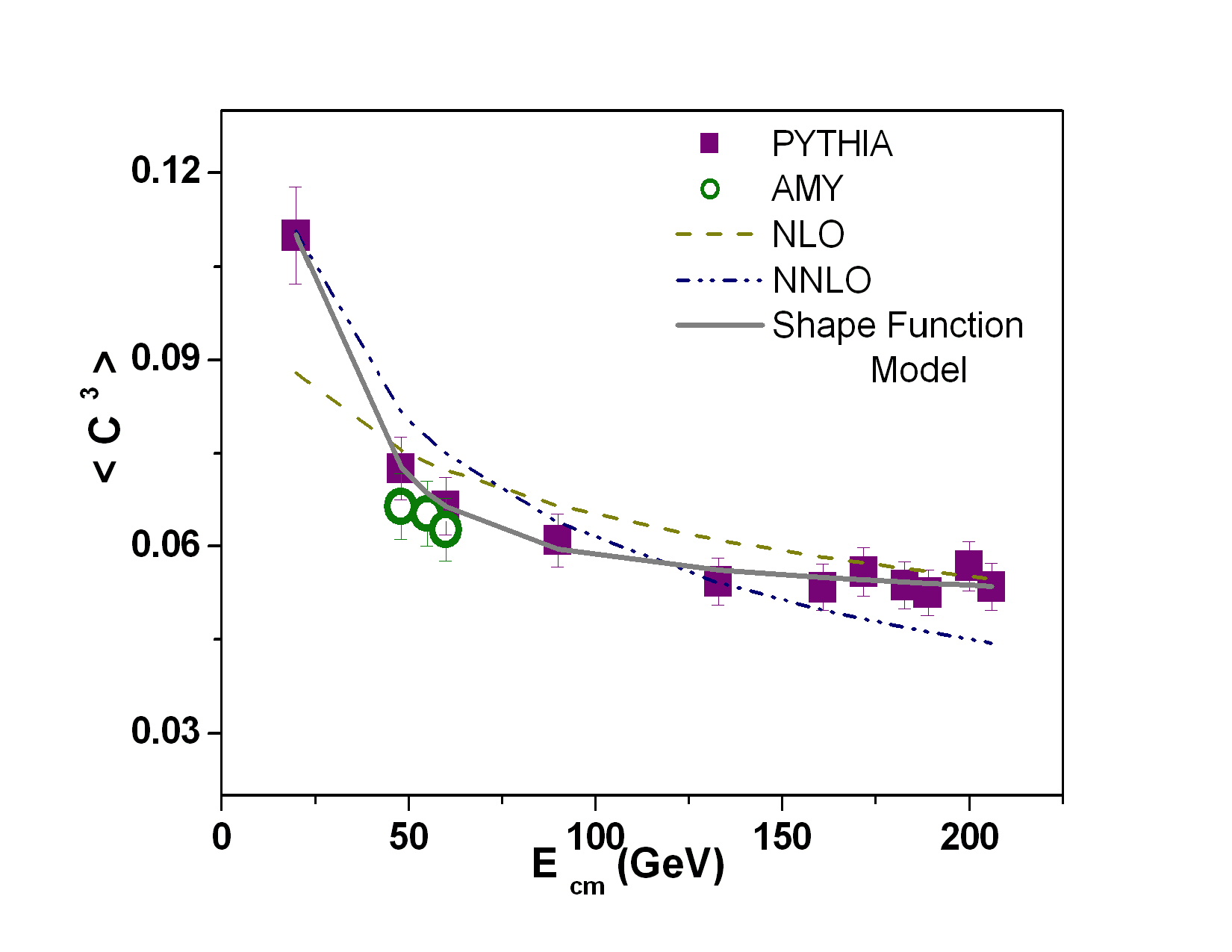}
\caption{\label{fig3} Fitting the C-parameter in shape function model with data. }
\end{figure}

Figure \ref{fig3} shows the distributions obtained from the experimental data as well as PYTHIA. By fitting the shape function model with the corresponding distribution, we obtain the strong coupling constant $\alpha_{s} (M_{Z^{0}})$  and the non-perturbative parameter $(\lambda_{1})$. Our results are indicated in Table \ref{tab4}.

\begin{table}[b]%The best place to locate the table environment is directly after its first reference in text
\caption{\label{tab4}%
Measurements of the coupling constants using the shape function model.}
\centering
\begin{tabular}{||c|c|c|c||}\hline
\textrm{\textbf{n}}&\textrm{\textbf{1}}&{\textrm{\textbf{2}}}&\textrm{\textbf{3}}\\
\hline
\textbf{$\alpha_{s}(M_{Z^{0}})$} & $0.149\pm 0.003$ & $0.143\pm 0.003$ & $0.140\pm 0.0064$ \\
\textbf{$\lambda_{1}$} & $0.578\pm 0.037$ & $0.521\pm 0.039$ & $0.449\pm 0.033$ \\ \hline
\end{tabular}
\end{table}
\clearpage
These values are in good agreement with the results from other event shape variables carried out in our previous publications \cite{ref3, ref4}. Our results are also consistent with the QCD predictions \cite{ref13}.

\subsection{\label{3-4}The variance}
The simple prediction for the variance of event shape variables (y) on hadron level is \cite{ref19}:
\begin{equation}\label{19}
Var(y)=\langle y^{2} \rangle - \langle y \rangle ^{2}.
\end{equation}

We can obtain a purely perturbative expression for the variance in the dispersive model and also in the shape function model, up to strongly suppressed corrections $O(\alpha_{S}⁄(Q^{2}))$. The first and second orders give the identical prediction in NLO and NNLO moments.

 However the NLO and NNLO do not include non- perturbative part. 

 Thus in case of the C-parameter, we have: 
\begin{equation}\label{20}
Var(C)=\langle C^{2} \rangle_{NLO} - \langle C \rangle ^{2}_{NLO}.
\end{equation}

\begin{figure}
   \centering
\includegraphics[scale=0.23]{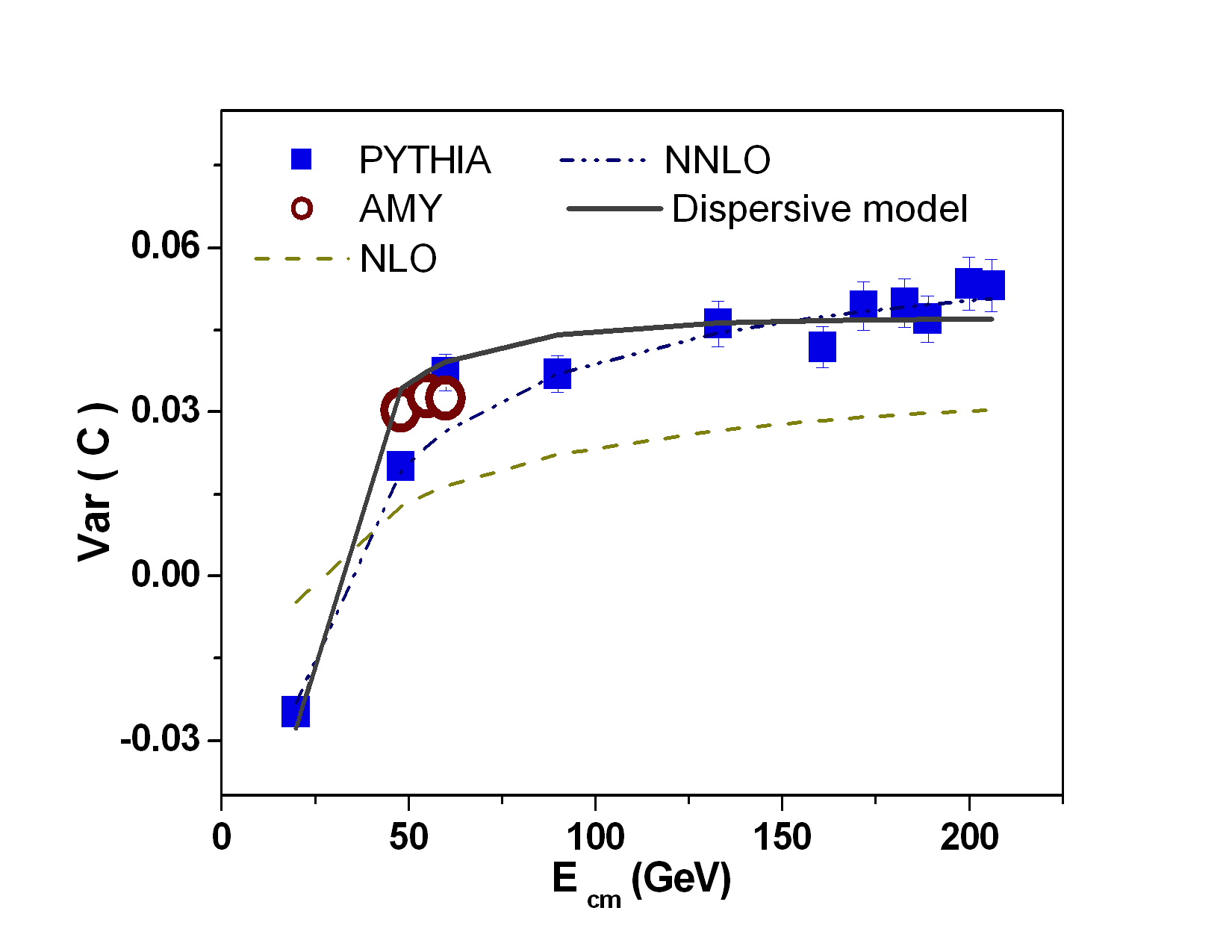}
\caption{\label{fig4} Fitting variance by using the dispersive model.}
\end{figure}

\begin{figure}
\includegraphics[scale=0.23]{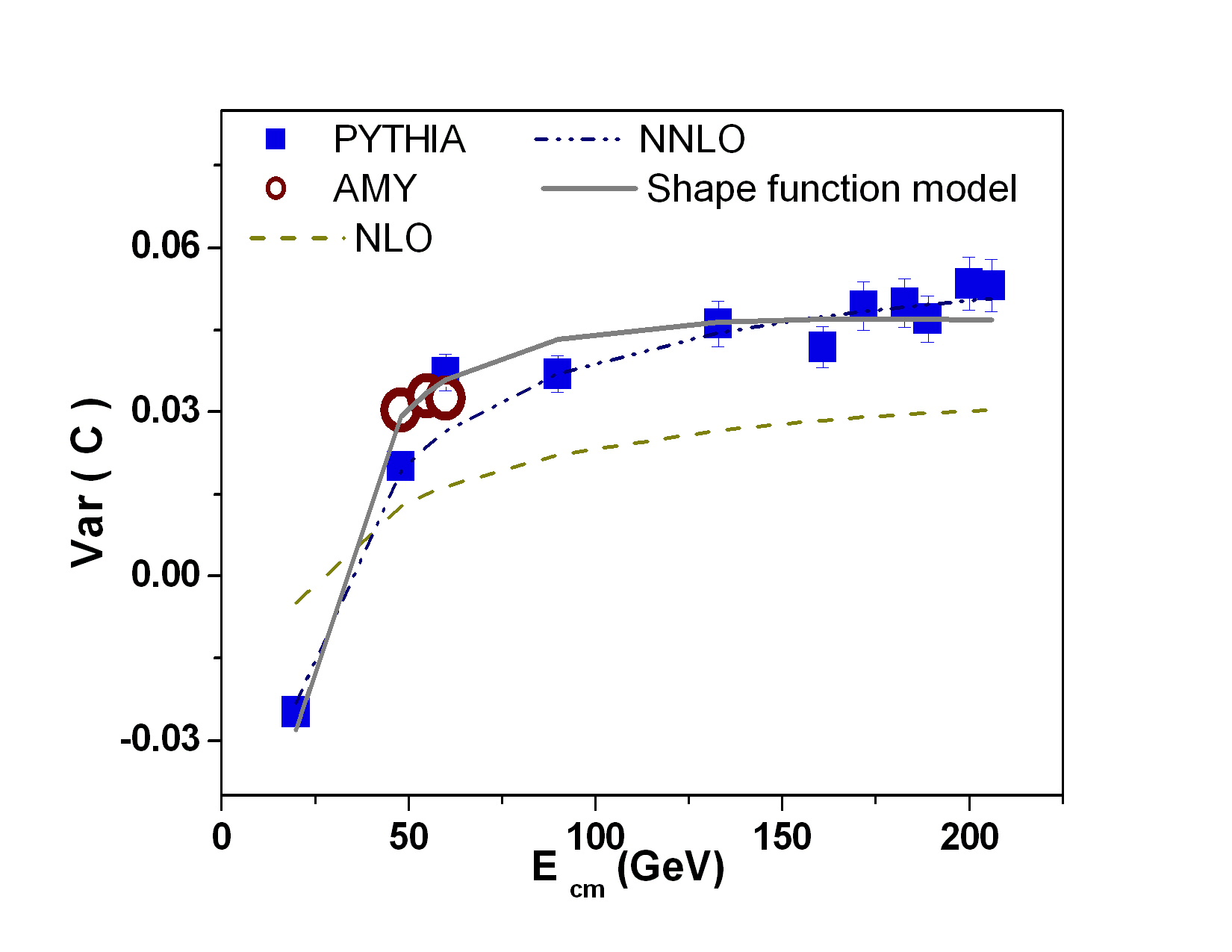}
\centering
\caption{\label{fig5} Fitting variance by using the shape function model.}
\end{figure}

On the other hand, if we take into account the non-perturbative part of the model, we will have:
\begin{equation}\label{21}
Var(C)=\langle C^{2} \rangle_{total} - \langle C \rangle ^{2}_{total}.
\end{equation}
where the subscript indicates  the  perturbative as well as the non- perturbative parts of our calculations in both models. In Figures \ref{fig4} and \ref{fig5} we show the variance for dispersive as well as shape function model. The obtained results by fitting variance with models are summarized in Table \ref{tab5} and Table \ref{tab6} respectively.

\begin{table}[b]%The best place to locate the table environment is directly after its first reference in text
\caption{\label{tab5}%
Measurements of the coupling constants by using the variance of dispersive model.}
\centering
\begin{tabular}{||c|c|c||}  \hline
\textbf{Observable}& \textbf{$\alpha_{s}(M_{Z^{0}})$} & \textbf{$\alpha_{0}(\mu_{I})$} \\ 
\hline
\textit{C-parameter} & $0.1039\pm 0.0059$ & $0.5183\pm 0.0211$ \\ \hline
\end{tabular}
\end{table}

\begin{table}[b]%The best place to locate the table environment is directly after its first reference in text 
\caption{\label{tab6}%
Measurements of coupling constants by using the variance of shape function model.}
\centering
\begin{tabular}{||c|c|c||} \hline 
\textbf{Observable}& \textbf{$\alpha_{s}(M_{Z^{0}})$} & \textbf{$\lambda_{1}$} \\ 
\hline
\textit{C-parameter} &$0.1052\pm 0.0031$ & $0.7164\pm 0.0342$ \\ \hline
\end{tabular}
\end{table}


Both tables indicate that our values for the coupling constant in perturbative and in the non-perturbative regions are in good agreement with the QCD predictions. They are also consistent with those obtained from other experiments \cite{ref13}.
\clearpage

‎

\end{document}